\documentclass{article}

\usepackage{graphicx}
\usepackage{fullpage,amsmath,amssymb,latexsym}
\usepackage{lineno}
\usepackage{multirow}
\usepackage{natbib}
\begin{document}
\title{Jupiter and Saturn Rotation Periods}
\author{Ravit Helled$^1$\footnote{corresponding author, rhelled@ucla.edu}, Gerald Schubert$^1$, and John D. Anderson$^2$\\
\small{$^1$Department of Earth and Space Sciences and Institute of Geophysics and Planetary Physics,}\\
\small{University of California, Los Angeles, CA 90095Ð1567, USA}\\
\small{$^2$Jet Propulsion Laboratory\footnote{Retiree},} \\
\small{California Institute of Technology, Pasadena, CA 91109}\\
}
\date{}
\maketitle \vskip 0cm \noindent

\begin{abstract}
Anderson \& Schubert (2007, Science,~317,~1384) proposed that Saturn's rotation period can be ascertained by minimizing the dynamic heights of the 100 mbar isosurface with respect to the geoid; they derived a rotation period of 10h 32m 35s. 
We investigate the same approach for Jupiter to see if the Jovian rotation period is predicted by minimizing the dynamical heights of its isobaric (1 bar pressure level) surface using zonal wind data. A rotation period of 9h 54m 29s is found. Further, we investigate the minimization method by fitting Pioneer and Voyager occultation radii for both Jupiter and Saturn. Rotation periods of 9h 55m 30s and 10h 32m 35s are found to minimize the dynamical heights for Jupiter and Saturn, respectively. Though there is no dynamical principle requiring the minimization of the dynamical heights of an isobaric surface, the successful application of the method to Jupiter lends support to its relevance for Saturn. \\
We derive Jupiter and Saturn rotation periods using equilibrium theory in which the solid-body rotation period (no winds) that gives the observed equatorial and polar radii at the 100 mbar level is found. Rotation periods of 9h 55m 20s and 10h 31m 49s are found for Jupiter and Saturn, respectively.  We show that both Jupiter's and Saturn's shapes can be derived using solid-body rotation, suggesting that zonal winds have a minor effect on the planetary shape for both planets.\\
The agreement in the values of Saturn's rotation period predicted by the different approaches supports the conclusion that the planet's period of rotation is about 10h 32m.  
\end{abstract}
\newpage

\section{Introduction}
Knowledge of a planet's rotation rate is fundamental to understanding its internal structure and atmospheric dynamics. The atmospheres of Jupiter and Saturn have strong zonal winds with equatorial speeds reaching $\sim$ 100 m s$^{-1}$ on Jupiter and $\sim$ 400 m s$^{-1}$ on Saturn. However, these zonal wind velocities are based on assumed values of the rotation periods of the planets, 9h 55m 29s  for Jupiter, and 10h 39m 22s for Saturn (Lindal, 1992). Jupiter's rotation period is assumed to be identical to the rotation period of its tilted magnetic field, which has not changed in many decades (Porco et al., 2003; Smith \& Hunt, 1976, Riddle \& Warwick, 1976; Higgins et al., 1996). The 10h 39m 22s rotation period of Saturn, on which the 400 m s$^{-1}$ equatorial zonal wind velocity is based, derives from the Voyager era periodicity in Saturn's kilometric radiation (Ingersoll \& Pollard, 1982; Dessler, 1983). That period is now known to be variable and Saturn's rotation rate is therefore uncertain (Gurnett et al., 2007). Accordingly, the atmospheric zonal wind velocities with respect to the underlying rotating planet are not known for Saturn.\\ 
Anderson \& Schubert (2007, hereafter paper I) used the improved gravitational data measured by Cassini (Jacobson et al., 2006) to infer the properties of Saturn's interior. Since the rotation rate of Saturn is unknown internal models were presented for rotation periods between 10h 32m 35s and 10h 41m 35s. The interior structure was modeled by applying the 'theory of figures' (Zharkov \& Trubitsyn, 1978) using a 6th order (in normalized radius) polynomial density function, assuming a solid-body rotation.  Anderson \& Schubert (2007) suggested that Saturn's rotation period could be constrained by minimizing the dynamical height variations with respect to the geoid of the 100 mbar isosurface of Saturn's atmosphere. The shape of Saturn's 100 mbar isosurface is known, but the shape of the geoid is not because it depends on Saturn's unknown rotation rate. A minimization of the dynamic heights of the 100 mbar isosurface was found for a rotation period of 10h 32m 35s. This rotation period enables computation of an interior model that best fits the measured gravitational moments, and is consistent with the lower bound rotation period from all available magnetic and SKR data (Gurnett et al., 2007). Figure 1 shows the latitude-dependent dynamic heights of the 100 mbar isobaric surface, measured relative to a reference geoid with rotation periods of 10h 32m 35s, 10h 35m 35s, and 10h 38m 35s (paper I). \par

Smith et al. (1982) commented that a more rapid Saturnian rotation rate (compared with the Voyager rate) would result in lower wind velocities with respect to the solid-body rotation, and that the dynamical heights are minimized for a rotation period between 10h 31m and 10h 32m. 
A shorter rotation period for Saturn's deep interior has also been suggested by Allison \& Stone (1983) who noted that a rotation period shorter than the Voyager radio rotation period by about 1\% could better explain the energetics of the Saturnian atmosphere. A shorter rotation period for the interior produces a Jovian-like Saturnian wind system with sensible equatorial wind speeds and both eastward and westward jets at higher latitudes. However, at that time, and during the last few decades there was no reason to question Voyager's rotation period, and the idea of a shorter rotation period for Saturn was not pursued. \par 
Cassini has recently measured the SKR period to be 10h 47m 6s (Gurnett et al., 2007), about 8 minutes longer than the Voyager period. In addition, during Cassini's orbit around Saturn this period was found to be changing with time (Gurnett et al., 2007). It is therefore clear that these measurements do not represent the rotation period of Saturn's deep interior.  The good fit of the gravitational data for an interior model with a faster rotation rate than the Voyager rate, as shown in paper I, and the realization that Saturn's rotation period is not given by the SKR period, call for new efforts to confirm the proposed faster rotation of Saturn. \par

In this paper we present different methods for inferring the rotation periods of the giant planets. Applying the  methods to Jupiter and  taking Jupiter's rotation period as unknown, we reconstruct Jupiter's SKR rotation period with an error of up to 1 minute.  We find that the dynamical heights minimization method (paper I) works successfully for Jupiter. However, there is no dynamical principle requiring the minimization of the dynamical heights of the isobaric surface and the method does not work for terrestrial planet atmospheres. The minimization method does work for Jupiter whose atmosphere has both eastward and westward jets. The successful application of the method to Jupiter lends support to  its relevance to Saturn, for which a rotation period of $\sim$10h 32m is derived. 

\section{Minimization of the Dynamical Heights of a Pressure Isosurface Using Wind Data}
Jupiter's physical shape, i.e., the radius of its isobaric surface (here taken as the 1 bar isosurface) as a function of latitude, can be represented by a combination of its reference geoid and the dynamical heights of the isobaric surface relative to the geoid. The reference geoid is the spheroid whose shape is determined by equilibrium theory for a given rotation period (Zharkov \& Trubitsyn, 1978). Using zonal wind data with respect to an assumed uniform rotation period, the radius of the corresponding potential surface as a function of latitude can be computed (Lindal et al., 1985; French et al., 1998). The isobaric surface does not coincide with the geoid because of the dynamical effects of zonal winds. The dynamical heights with respect to the reference geoid are given by (Lindal et al., 1985)
 \begin{equation}
h(\phi) = \frac{1}{g}\int_{\phi}^{\frac{\pi}{2}}V_W\left(2\omega_{\mathrm{ref}}+\frac{V_W}{r_{\mathrm{ref}}(\phi)cos\phi}\right)\frac{sin(\phi+\psi_{\mathrm{ref}})}{cos\psi_{\mathrm{ref}}}r_{\mathrm{ref}}(\phi)d\phi
\end{equation}
where $g$ is the average of the magnitude of the acceleration of the gravity (the vector acceleration of gravity varies both radially and latitudinally along a local vertical between the isobaric surface and the reference geoid), $\psi$ is a small angle that gives the difference between the planetocentric latitude $\phi$ and the planetographic latitude $\phi'$, $\psi =\phi'-\phi$, $r$ is the distance to the center of the planet, and $V_W$ is the zonal wind velocity. Parameters with the subscript 'ref' represent quantities calculated for the reference geoid (see Lindal et al.,~1985 for further details). For simplicity, and because vertical profiles of zonal wind speeds are unknown, the measured zonal wind velocities are taken to be independent of altitude (radial component). \par

For $V_W$ we use zonal wind data obtained from Hubble Space Telescope images during 1995 to 2000 (Garc\'{i}a-Melendo \& S\'{a}nchez-Lavega, 2001) at three different wavelengths (410,  892, 953 nm) representative of Jupiter's cloud top levels. Figure 2 shows the wind velocities for Jupiter as computed by Garc\'{i}a-Melendo \& S\'{a}nchez-Lavega (2001). These wind data are found to be in good agreement with previous zonal wind measurements for Jupiter (Limaye, 1986).\par

We calculate Jupiter's reference geoid for its measured rotation period of 9h 55m 29.7s (Lindal, 1992). The reference geoid is defined as the level surface of equal effective potential
\begin{equation}
U = \frac{G M}{r} \left( 1 - \sum_{n=1}^\infty \left( \frac{a}{r} \right)^{2 n} J_{2 n} P_{2 n} \left( \cos \theta  \right)  \right)+ \frac{1}{2} \omega^2 r^2 \sin^2 \theta.
\label{U}
\end{equation}
The level surface of constant effective potential $U$ is computed to the fifth order in the smallness parameter $q=\frac{\omega^2 a^3}{GM}$, where $a$ is the equatorial radius of the geoid, $GM$ is its mass multiplied by the gravitational constant, $\theta$ is the co-latitude, and $\omega$ is the angular velocity given by $2\pi/P$, with $P$ the rotation period. 
The polar radius is fixed at the value 66,854 km, the polar radius of the 1 bar isosurface. The value of the polar radius is taken as the average of the two polar radii at that pressure level. The northern and southern polar radii differ from each other due to the difference in the zonal wind strength at the two hemispheres (e.g., Ingersoll, 1970).
The harmonic coefficients $J_{2n}$ are obtained from the measured values for a reference equatorial radius of 71,492 km (Jacobson, 2003). The values of $J_{2n}$ from Jacobson (2003) in units of $10^{-6}$ are $J_2=14696.43\pm0.21$, $J_4=-587.14\pm1.68$, $J_6=34.25\pm5.22$. \par

The measured gravitational coefficients $J_{2n}$ at a reference radius of 71,492 km are of order $q^n$, and can be replaced in the effective potential function by $a^{2n}j_{2n}$=(71,492 km)$^{2n}J_{2n}$(measured). Since the equatorial radius $a$ of the reference geoid depends on the rotation period, so do the values of $j_{2n}$. 

The radius $r$ as a function of $cos\theta$ can be expanded by 
\begin{equation}
r(cos\theta)= a\left( 1+ \sum_{n=1}^{\infty}r_n(cos\theta) q^n\right).
\end{equation}
To find the reference geoid we substitute the expression for $r$ (equation (3)) into equation (2) and expand the result in the smallness parameter q. 
Then a zero function can be defined by,
\begin{equation}
F(cos\theta)= U(cos\theta) -U(0)
\end{equation}
where, to the fifth order in $q$,
\begin{equation}
U(0) = 1+ \frac{1}{2}q + \frac{1}{2}j_2q-\frac{3}{8}j_4q^2+\frac{5}{16}j_6q^3-\frac{35}{128}j_8q^4+\frac{63}{256}j_{10}q^5.
\end{equation}
The coefficients $r_n$ are then found recursively by setting the coefficients of $q$ in the expanded equation (4) to zero, starting with order one. 
The results for the first three $r_n$ functions are,
\begin{eqnarray}
r_1 &=& -\frac{1}{2} \left( 1 + 3 j_2  \right) cos\theta^2 \nonumber \\
r_2 &=& - \frac{1}{4} \left( 2 + 3 j_2 - 9 j_2^2 - 15 j_4  \right)cos\theta^2 + \frac{1}{8} \left( 6 + 6 \left( 1 - 6 j_2  \right) j_2 - 35 j_4  \right) cos\theta^4 \nonumber \\
r_3 &=& - \frac{1}{16} \left( 8 - 27 j_2 \left( 2 \left( 1 - j_2 \right)j_2 -5 j_4       \right) - 45 j_4 + 105 j_6      \right) cos\theta^2 \nonumber \\
&& + \frac{5}{16} \left(6 -8 j_4 + 3 j_2 \left( 2 + 6 j_2 \left( 3 j_2 -1 \right) + 43 j_4 \right)  +63 j_6   \right) cos\theta^4  \nonumber \\
&& - \frac{1}{16} \left(24 + 35 j_4 + 18 j_2 \left( 2 + j_2 + 21 j_2^2 +35 j_4 \right) + 231 j_6 \right) cos\theta^6
\label{rn}
\end{eqnarray}
The expressions are considerably more complicated for orders 4 and 5, but they can be evaluated similarly. The point in going to fifth order is not to introduce measured values of $J_8$ and J$_{10}$, they have not been detected so far; fifth order is used for purposes of representing the measured radii to a precision commensurate with their accuracy at the $\sim$~2~km level.\par

This procedure provides an equation for $a/r$. We then determine $a/b$ where $b$ is the polar radius, by $a/b=a/r(cos\theta=0)$. Since the polar radius is assumed to be known (66,854 km), we finally find the equatorial radius $a$, and the values of $j_{2n}$ for an assumed rotation period $\omega$ (see paper I for further details). \par

Once the reference geoid is calculated, Jupiter's physical shape can be derived. The physical shape represents the height of the 1 bar level with respect to the planet's center, and is constructed by adding Jupiter's dynamical heights $h(\phi)$ to 
the reference geoid corresponding to the measured rotation period of  9h 55m 29.7s. We can now search for the rotation period which minimizes the dynamical heights of the isobaric surface (equation (1)). To do so, we construct reference geoids for four rotation periods: 9h 57m 29.7s, 9h 55m 29.7s, 9h 54m 29.7s, and 9h 53m 29.7s. 
The altitudes of the 1 bar isosurface  above the reference geoids (dynamical heights $h(\phi)$) for different assumed rotation periods are found by subtracting the calculated geoids (with varying rotation periods) from the physical shape of Jupiter (determined from the dynamic heights with respect to the 9h 55m 29.7s geoid). \par 

Table 1 provides the physical parameters of the geoid for the four assumed rotation periods. 
The amplitudes of the 1 bar isosurface above the reference geoid for the four different rotation periods are shown in Figure 3. The dynamical heights for Jupiter's measured rotation period are found to be in agreement with the ones calculated by Ingersoll (1970).
The rotation periods investigated are 1 minute shorter and 2 minutes shorter and longer than the actual measured rotation period of Jupiter, because the inherent standard error in rotation period using the dynamical height minimization method was found to be about $\pm1$ minute, for both Jupiter and Saturn. The standard error for the wind radii is a few kilometers resulting in an error in rotation period of up to 1 minute (paper I).
As can be seen from the figure the dynamical heights are minimized for a rotation period of 9h 54m 29.7s, a rotation period which is only one minute faster than Jupiter's measured rotation period. For Saturn, the method results in a rotation period of 10h 32m 35s (paper I). 

\subsection{Minimizing the Dynamical Height Difference Between the Equator and the Pole}
A simplified version of the dynamical height minimization method presented above minimizes the dynamical height difference between the planet's pole and equator. 
We take $v_{V}$ to be the measured cloud-tracked wind velocity, referenced to the Voyager radio period  (Garc\'{i}a-Melendo \& S\'{a}nchez-Lavega, 2001). The zonal wind measured with respect to a 'revised reference geoid' $v_{r}$ varies with the revised planetary rotation period as (Allison \& Stone, 1983)
\begin{equation}
v_{r} = v_{V} + 2\pi a\left(\frac{1}{\tau_{V}}-\frac{1}{\tau_{r}}\right)cos\phi
\end{equation}
where $a$ is the planet's equatorial radius, $\tau_{V}$ is the Voyager rotation period, and $\tau_{r}$ is the revised planetary rotation period.\par

The dynamical heights with respect to the reference geoid (equation (1)) for small wind velocities and negligible small non-geostrophic contributions to the integral are given by (Smith et al., 1982; Lindal et al., 1985),
\begin{equation}
h(\phi) = \frac{2 \omega_{V} a}{\bar{g}}\int_{\phi}^{\frac{\pi}{2}}v_V sin\phi d\phi ,
\end{equation}
where $\omega_{V}$ is the Voyager angular velocity and $\bar{g}$ is the average acceleration of gravity.
Then, by substituting equation (7) for the zonal wind into equation (8), 
the implied change in the dynamical height difference between the equator and the pole can be written as
\begin{equation}
h_{r} = h_{V} + 2\pi\omega_{V}\frac{a^2}{\bar{g}}\left(\frac{1}{\tau_{V}}-\frac{1}{\tau_{r}}\right),
\end{equation}
where $h_V$ is the Voyager dynamical height. 
Assuming that the dynamical heights are minimized ($h_{r}$ goes towards zero), equation (9) can be solved for $\delta\tau = \tau_{r} - \tau_{V}$ with the result 
\begin{equation}
\delta\tau = -\tau_{V} \frac{\bar{g}h_{V}}{(\omega_{V}a)^2}.
\end{equation}
Substituting the values for Jupiter ($h_{V}$ $\sim$ 12 km, $\bar{g}\sim$ 24 m s$^{-2}$, $(\omega_{V}a)^2 \sim$ 1.58$\times10^8$ m$^2$ s$^{-2}$ , and $\tau_{V}=$ 9.925h = 9h 55m 30s), we get  
$\delta \tau \sim$ -1 minute, essentially the same as the 1 minute shorter period
derived earlier. \par 

Applying this approach to Saturn, with $h_{V}$ = 120 km, $\bar{g}\sim$ 10 m s$^{-2}$, $(\omega_{V}a)^2 = 9.765 \times10^7$ m$^2$  s$^{-2}$, and
$\tau_{V}$ = 10.6567h = 10h 39m 24s (Lindal et al., 1985), we get $\delta \tau \sim$ -7 minutes, a value in excellent agreement with the revised period of 10h 32m 35s derived in paper I. Although the uncertainty of these estimates is a few tens of
seconds, we show that the simplified version of the dynamical height minimization method produces comparable results to the ones obtain above.

\section{Fit to Pioneer and Voyager Occultation Radii}
The physical shape of Jupiter and Saturn can be derived independently of zonal wind data by using occultation radii. 
While in the previous section the physical shape was derived from the geostrophic equation, occultation radii provide the planetary shape independently of wind data.  \par

Six radio occultation radii of Jupiter's 100 mbar isosurface from the Pioneer and Voyager missions are given
in Fig. 7 of Lindal et al. (1981). Using fifth-order expressions for the radius (equation (3)), we recreate the geoid of Lindal et al. (1981) by using the same  gravitational field (Null, 1976) as used in their paper. The equatorial radius of
this geoid is 71,541 km at the 100 mbar level, with a polar radius
of 66,896 km. 
To construct Jupiter's physical shape (occultation radii at their corresponding geocentric latitudes) independently of the underlying geoid, we add the heights of the 100 mbar isosurface with respect to the geoid by digitizing the right hand panel of Fig.7 in Lindal et al. (1981) to two places past the decimal, well inside their error bars of $\sim$ 2 km.   
\par

We next search for a rotation period that can best fit the occultation radii (physical shape), i.e., a period that minimizes the heights between the calculated geoid (for varying rotation periods) and the occultation radii. 
We introduce three free parameters into the fitting function. The first is the equatorial radius $a$ at the 100 mbar level, the second is the period of rotation $P$, and the third is an offset $\Delta Z$ of the center of mass along the polar axis, as suggested by Lindal et al. (1981). This offset produces a better fit to the occultation radii. Both $a$ and $P$ are used to update the smallness parameter $q$, and the harmonic coefficients $j_{2n}$. The computed values of the six radii for each geoid (rotation period) follow from equation (3), and the offset along $z$ is then added to the calculated radii. \par

The residuals in radii at 100 mbar (the measured occultation radii minus the computed radii derived from the geoid) are collected into a six dimensional column matrix $y$, and the linear system $y = A x$ is inverted to obtain corrections to the assumed values of the three parameters. The inverse of the $6 \times 3$ rectangular matrix $A$ is the pseudo inverse in the singular value decomposition (SVD) method (Lawson \& Hanson, 1974). The process is repeated to convergence. The matrix A is the design matrix in the least squares process. It is computed by applying small finite differences to the three parameters and by evaluating the linear effect on the residuals. The resulting $A$ matrix is well conditioned and the least squares solution is full rank three. The converged covariance matrix is $(A^T A)^{-1} \sigma_r^2$, with $T$ indicating a transpose. The variance $\sigma_r^2$ is estimated by $(y^T y)/3$ with the result $\sigma_r = 2.38$ km. 
Using this approach one does not need to set the polar radius a priori. Once the solution for the equatorial radius is found, the value of the polar radius $b$ is obtained from equation (3) with $cos\theta$ set to one.   
\par 

A best fit to the occultation radii is found for a rotation period of $9~{\rm h}~55~{\rm m}~27~{\rm s} \pm 15~{\rm s}$, a rotation period in excellent agreement with Jupiter's measured rotation period. The equatorial radius $a$, polar radius $b$, and the offset $\Delta Z$ for this rotation period are found to be $71,540.8~{\rm km} \pm 1.4~{\rm km}$, 66,895.2 km $\pm$ 3.9 km, and $5.9~{\rm km} \pm 1.7~{\rm km}$, respectively. It should be noted that the errors given are formal mathematical estimates and that they do not reflect the real uncertainty in the rotation period estimate which can be of the order of about 1 minute. \par

The planetocentric latitude
$\phi$ and the radii $r$ of the 100 mbar isosurface are listed in Table 2.
We also do an extrapolation of the radii to the one-bar level, for purposes of
comparing with the elevations inferred from the wind data to the 1 bar level. This extrapolation
is accomplished by subtracting 47.82 km along the vertical to the geoid and
projecting it on the geocentric radius. The estimate of the 47.82 km height
of the 100 mbar level above the one-bar level in the atmosphere is obtained
by interpolating in Table 8.2 in Lodders \& Fegley (1998). We get fairly good agreement with the 1 bar isosurface (1 bar shape) derived in the previous section based on the wind data. The fourth column ('Residuals') presents the height difference between the calculated geoid for the 100 mbar pressure level (with the offset $\Delta Z$ included) and the measured occultation radii for the best fit model (P = 9h 55m 27s). For our best fit  rotation period, the residuals do not exceed 3 km. As in previous sections, but using another source for Jupiter's physical shape, we find that the rotation period of Jupiter's deep interior can be ascertained by minimizing the dynamical heights of the (100 mbar) isosurface. 
\par

Both radio occultation radii and wind radii for Saturn have been fit in paper I for four rotation periods from 10h 32m 35s to 10h 41m 35s. The approach used here for Jupiter can be applied to the Saturn occultation data. As for Jupiter, this calculation uses a physical shape which is independent of wind data. 
Unfortunately, the Pioneer and Voyager occultation radii are not well distributed, with four of the five latitudes being in the southern hemisphere. As far as we know, there are no published radii from Cassini radio occultations. For Saturn we set the Z offset at -5.75 km and search for the rotation period and equatorial radius that can best reconstruct the occultation radii. The rotation which best fits the occultation radii is found to be 10h 32m 35s $\pm$ 17s, essentially, the same period as found in paper I. 
The equatorial radius $a$ and  polar radius $b$ for this rotation period are found to be 60,355.4 km $\pm$ 2.9 km and 54,438 km $\pm$ 10 km, respectively.  The five occultation radii and their residuals for the best fit model (P = 10h 32m 35s) are shown in Table 3. The larger residual values for Saturn, as compared to Jupiter, may be a result of more dynamic horizontal wind motions for Saturn, and hence a larger deviation of the isobaric surface from the geoid surface. As a result, the uncertainty in Saturn's rotation period might be larger than that of Jupiter and could be about a couple of minutes.    

\subsection{Jupiter and Saturn Rotation Periods from Equilibrium Theory}
If one uses the measured (independent of rotation period) polar and equatorial radii of a planet, the rotation period which best reproduces the planetary shape can be found using equilibrium theory (Zharkov \& Trubitsyn, 1978). The measured radii of Jupiter and Saturn are given in table 4. This treatment of the effective potential assumes a negligible departure from solid-body rotation within the atmosphere and interior {and therefore assumes that the effect of the winds on the planetary shape is minimal}. In principle, accounting for differential rotation can influence the planetary figure as well as the gravitational harmonics (Hubbard, 1982; 1999). However, as we will show below, the shapes of Jupiter and Saturn are found to be accurately reproduced using a solid-body rotation.  \par

The effective potential of the planet (equation (2)) can be used to determine the potential at both the pole and the equator. By equating the two results for the potential, the rotation rate of the planet can be found. 
To third order the rotation period is given by (Zharkov \& Trubitsyn, 1978),
\begin{equation}
\tau_{rot} = 2\pi \sqrt{\frac{r_e^3}{2GM}}\left(\left(\frac{r_e}{r_p}\right)-1-J_2\left[\left(\frac{1}{2}\right) + \left(\frac{r_e}{r_p}\right)^3 \right] -J_4\left[\left(-\frac{3}{8}\right) + \left(\frac{r_e}{r_p}\right)^5 \right] -J_6\left[\left(-\frac{5}{16}\right) + \left(\frac{r_e}{r_p}\right)^7 \right]\right)^{-1/2}
\end{equation}
where $r_e$ and $r_p$ are the equatorial and polar radii, respectively. \par 

The equatorial radius used should be consistent with the reference equatorial radius used for the calculation of the gravitational coefficients. For Jupiter we use Jacobson's (2003) reference equatorial radius and gravitational harmonics as listed in Table 4. Both the equatorial and polar radii correspond to the 1 bar pressure level. 
Jupiter's rotation period is found to be 9h 55m 20s, only 10 seconds shorter than Jupiter's measured radio period. This agreement can be taken as confirmation that the dynamical deformation of Jupiter's figure by its differential wind motion is negligible compared with the equatorial bulging due to the planet's basic rotation. We repeat this exercise using values of the 100 mbar pressure level, for which Jupiter's equatorial and polar radii are  71,541 km and 66,896 km, respectively (Lindal et al., 1981). The adjusted gravitational coefficients for Jupiter at the 100 mbar pressure level and its radii are listed in table 4. 
Using the values for the 100 mbar pressure level a rotation period of 9h 55m 20s is derived, identical to the one derived for the 1 bar pressure level.  \par

For Saturn we use the gravitational field measured by Cassini (Jacobson et al., 2006). These gravity data however, correspond to a reference equatorial radius of 60,330 km and not to Saturn's equatorial radius of  60,268 km at the 1 bar pressure level . We therefore adjust the gravitational harmonics to fit an equatorial radius of 60,268km and a polar radius of 54,364 km (see table 4). We find that Saturn's shape is reproduced for a rotation period of 10h 31m 49s. This rotation period is less than one minute shorter than the rotation period derived by the dynamical heights minimization method. For the 100 mbar pressure level we take the equatorial and polar radii to be 60,367 km and 54,438 km, respectively (Lindal et al., 1985). The corresponding gravitational coefficients are given in table 4. The rotation period which best reproduces Saturn's shape at the 100 mbar pressure level is found to be 10h 31m 50s,  a period in excellent agreement with the value derived for the shape at the 1 bar pressure level. \par
 
\section{Summary and Conclusions}
Our work suggests that the rotation period of Jupiter's deep interior can be obtained by minimizing the dynamical heights of its wind level pressure isosurface. We find that a rotation period of 9h 54m 29.7s $\pm 1$m minimizes the dynamical heights of the 1 bar isosurface. A simpler version of this method minimizes the dynamical height difference between the equator and the pole (section 2.1). 
This procedure provides the same rotation period. 
The derived rotation period, however, is based on the dynamical height calculation which is valid for a vertically hydrostatic, pressure-balanced flow, and assumes a vertically uniform velocity. Measurement errors may lead to discrepancies as well. 
As a result, the calculation of the dynamical heights can be in error by a few kilometers, resulting in an error of about 1 minute in rotation period. Thus, the results suggest that Jupiter's rotation period can be derived by minimizing the dynamical heights of its isobaric surface, as suggested in paper I for Saturn. \par 

We investigate another way of examining the dynamical heights minimization by fitting Pioneer and Voyager radio occultation radii at the 100 mbar
level in the atmosphere. Again, we search for a rotation period which minimizes Jupiter's (100 mbar) isosurface, and obtain a best fit rotation period of 9h 55m 27s, consistent with Jupiter's IAU system III rotation of 9h 55m 29.7s.
We also use Pioneer and Voyager radio occultation radii at the 100 mbar level for Saturn, and search for the rotation period that minimizes the dynamical heights. A rotation period of 10h 32m 35s is derived, identical to the rotation period derived in paper I. \par

The successful application of the dynamical height minimization method to Jupiter lends support to its relevance for Saturn. We therefore conclude that the rotation period of Saturn's deep interior is about 10h 32m.
Although we conclude that Jupiter and Saturn rotation periods can be derived by minimizing the dynamical heights of their pressure isosurfaces, it is not clear whether the same method can be applied to Uranus and Neptune. An analysis of the dynamical height minimization method for Uranus and Neptune is more challenging due to larger uncertainties in the planetary physical parameters (radii, gravitational moments, rotation periods, etc.) as well as lack of accurate wind measurements and published occultation radii. \par

Yet another approach for deriving the rotation periods of Jupiter and Saturn uses equilibrium theory. Rotation periods of 9h 55m 20s and 10h 31m 49s are found for Jupiter and Saturn, respectively. \par

The faster rotation of Saturn advocated here is also supported by an independent analysis of the dynamical state of Saturn's atmosphere by Read et al.~(2009). These authors analyze the latitudinal distribution of potential vorticity on Saturn and show that at many latitudes the atmosphere is close to a state of marginal stability with respect to Arnol'd's second stability theorem, an atmospheric configuration allowing long planetary Rossby waves that reach deep onto the interior to be coherent across adjacent alternating jet streams. Read et al.~(2009) infer a rotation period of Saturn equal to 10h 34m 13s $\pm$ 20s, about one and a half minutes longer than the  $\sim$ 10h 32m 35s period proposed here. The Read et al.~(2009) analysis provides a dynamical basis for the faster rotation of Saturn. 

Finally, we find that the dynamical deformation of Jupiter by differential wind motions is negligible. By assuming the same for Saturn, we infer a rotation period of 10h 32m. We suggest that the Jovian and Saturnian figures are dominated by the planetary rotation and that the zonal winds have a minor effect on the  shapes of both planets.\par 

The successful inference of Jupiter's known rotation period and the agreement in the values of Saturn's rotation period, using different approaches, support the conclusion that Saturn's rotation rate is faster than previously thought. 

\subsection*{Acknowledgments} 
The authors thank Andy Ingersoll and an anonymous referee for valuable comments and suggestions.
R. H. and J. D. A acknowledge support from NASA through the Southwest Research Institute. G. S. acknowledges support from the NASA PGG and PA programs.

\section{References}
Allison, M. \& Stone, P. H., 1983. Saturn meteorology - A diagnostic assessment of thin-layer configurations for the zonal flow. 
Icarus, 54, 296--308.\\
Anderson, J. D., \& Schubert, G. 2007, SaturnÕs Gravitational Field, Internal Rotation, and Interior Structure. Science, 317, 1384--1387, (paper I).\\
Dessler, A. J., 1983, Physics of the Jovian magnetosphere. Cambridge University Press, New York, 498\\ 
French, R. G., McGhee, C. A. \& Sicardy, B., 1998. NeptuneÕs Stratospheric Winds from Three Central Flash Occultations. Icarus, 136, 27--49.\\
Garc\'{i}a-Melendo \& S\'{a}nchez-Lavega, 2001, A Study of the Stability of Jovian Zonal Winds from HST Images: 1995-2000. Icarus, 152, 316--330.\\
Gurnett, D. A., Persoon, A. M., Kurth, W. S., Groene, J. B., Averkamp, T. F., Dougherty, M. K., \&
Southwood, D. J. 2007, The Variable Rotation Period of the Inner Region of SaturnÕs Plasma Disk.  Science, 316, 442--445.\\
Higgins, C. A., Carr, T. D. \& Reyes, F. 1996. Geophys. Res. Lett., 23, 2653--2656\\
Hubbard, W. B., 1982. Effects of Differential Rotation on the Gravitational Figures of Jupiter and Saturn. Icarus, 52, 509--515.\\
Hubbard, W. B., 1999, NOTE: Gravitational Signature of Jupiter's Deep Zonal Flows. Icarus, 137, 357--359.\\    
Ingersoll, A. P., 1970. Motions in Planetary Atmospheres and the Interpretation of Radio Occultation Data. Icarus, 13, 34\\
Ingersoll, A. P. \& Pollard, D., 1982, Motion in the interiors and atmospheres of Jupiter and Saturn - Scale analysis, anelastic equations, barotropic stability criterion. Icarus, 52, 62--80.\\
Jacobson, R. A. 2003, JUP230 orbit solution, http://ssd.jpl.nasa.gov/?gravity\_fields\_op \\
Jacobson, R. A., Antreasian, P. G., Bordi, J. J., Criddle, K. E., Ionasescu, R., Jones, J. B., Mackenzie, R. A.,
Meek, M. C., Parcher, D., Pelletier, F. J., Owen, Jr., W. M., Roth, D. C., Roundhill, I. M., \& Stauch,
J. R. 2006, The Gravity Field of the Saturnian System from Satellite Observations and Spacecraft Tracking Data. ApJ, 132, 2520--2526\\
Lawson, C. L. \& Hanson, R. J., 1974. Solving least squares problems. Prentice-Hall
Series in Automatic Computation, Englewood Cliffs: Prentice-Hall, 1974.\\
Lindal, G. F. 1992, The atmosphere of Neptune - an analysis of radio occultation data acquired with Voyager 2. AJ, 103, 967--982.\\
Lindal, G. F., Sweetnam, D. N., \& Eshleman, V. R. 1985, The atmosphere of Saturn - an analysis of the Voyager radio occultation measurements. AJ, 90, 1136--1146.\\
Lindal, G. F., Wood, G. E. , Levy, G. S., Anderson, J. D. , Sweetnam, D. N., 
Hotz, H. B., Buckles, B. J., Holmes, D. P. , Doms, P. E. , Eshleman, V. R., 
Tyler, G. L. \& Croft, T. A., 1981. The atmosphere of Jupiter - an analysis of the
Voyager radio occultation measurements. J. Geophys. Res., 86, 8721--8727\\
Limaye, S. S. 1986, Jupiter: New estimates of the mean zonal flow at the cloud level. Icarus 65, 335--352.\\
Lodders, K. \& Fegley, B., 1998. The planetary scientistÕs companion / Katharina
Lodders, Bruce Fegley. The planetary scientistÕs companion / Katharina
Lodders, Bruce Fegley. New York : Oxford University Press, 1998. QB601.L84 1998.\\
Null, G. W., 1976. Gravity field of Jupiter and its satellites from Pioneer 10 and
Pioneer 11 tracking data. ApJ, 81:1153Ð1161, December 1976.\\
Porco, C. C., West, R. A., McEwen, A., Del Genio, A. D.; Ingersoll, A. P. et al. 2003, Science, 299, 154\\
Read, P. L., Dowling, T. E. \& Schubert, G., 2009. Rotation periods of Jupiter and Saturn from their atmospheric planetary-wave configurations. Nature, in press. \\ 
Riddle, A. C. \& Warwick, J. W. 1976, Redefinition of System III longitude, Icarus, 27, 457--459\\
Smith, B. A. \& G. E. Hunt, 1976, Jupiter University of Arizona Press, Tucson, 564\\ 
Smith, B. A., Soderblom, L., Batson, R., Bridges, P., Inge, J., Masursky, H., Shoemaker, E., Beebe, R., Boyce, J., Briggs, G., Bunker, A., Collins, S. A., Hansen, C. J., Johnson, T. V., Mitchell, J. L., Terrile, R. J., Cook, A. F., Cuzzi, J., Pollack, J. B., Danielson, G. E., Ingersoll, A., Davies, M. E., Hunt, G. E., Morrison, D., Owen, T., Sagan, C., Veverka, J., Strom, R. \& Suomi, V. E. 1982. A new look at the Saturn system: The Voyager 2 images. Science,  215, 505--537\\
Zharkov, V. N., \& Trubitsyn, V. P. 1978, Physics of planetary interiors (Astronomy and Astrophysics Series,
Tucson: Pachart, 1978) 
\clearpage

\begin{table}[h!]
\begin{center}
{\renewcommand{\arraystretch}{0.6}
\vskip 8pt
\begin{tabular}{l c c c c}

Jupiter Rotation Period & 9h 57m 29.7s & 9h 55m 29.7s & 9h 54m 29.7s & 9h 53m 29.7s
\\[1pt]
\hline
$a$ (km) & 71468.7 & 71491.5 & 71503. & 71514.5
\\[1pt]
$q$ & 0.0885127 & 0.0891935 & 0.089537 & 0.0898826
\\[1pt]
$J_2$ (10$^{-6}$) & 14706.0 & 14696.6 & 14691.9 & 14687.2
\\[1pt]
$J_4$ (10$^{-6}$) &-587.906 & -587.157 & -586.8 & -586.4
\\[1pt]
$J_6$ (10$^{-6}$) & 34.317 & 34.3 & 34.2& 34.2
\\[1pt]
\hline
\end{tabular} 
}
\caption{\label{geoid} 
Physical parameters for reference geoids with rotation periods between 9h 57m 29.7s and 9h 53m 29.7s.}
\end{center} 
\end{table}

\begin{table}[h!]
\begin{center}
\begin{tabular}{lcccl}
\multicolumn{5}{c}{\bf Occultation Radii for Jupiter} \\
\hline
\hline
$\phi$ (deg) & $r$, 100 mbar  & $r$, one bar  & Residuals & Spacecraft \\
& (km) & (km) & (km) & \\
\hline
60.3 & 67933.93 & 67886.18 & -0.94  & Pio 10 \\
28.0 & 70415.08 & 70367.34 &  2.97 & Pio 10 \\
19.8 & 70943.95 & 70896.17 & -0.98 & Pio 11 \\
0.07 & 71538.61 & 71490.79 & -2.23 & Voy 1 \\
-10.1 & 71378.73 & 71330.92 & 1.13 & Voy 1 \\
-71.8 & 67293.64 & 67245.85 & 0.04 & Voy 2 \\
\hline
\end{tabular}
\caption[Hello]{
	\label{radii}
	Jupiter's planetocentric latitude $\phi$ and radius $r$ obtained by a digitization of Fig. 7 in Lindal
et al. (1981) as described in the text. The radii at one bar are extrapolated from the measured 
radii at 100 mbar under the assumption of no vertical wind gradients. 
The residuals are for the 100 mbar radii and are from a converged fit with the period as a free parameter. The best fit model is found for a rotation period of 9h 55m 27s. }
\end{center}
\end{table}


\begin{table}[h!]
\begin{center}
\begin{tabular}{lcccl}
\multicolumn{5}{c}{\bf Occultation Radii for Saturn} \\
\hline
\hline
$\phi$ (deg) & $r$, 100 mbar  & $r$, one bar  & Residuals & Spacecraft \\
& (km) & (km) & (km) & \\
\hline
30.5 & 58545.4 & 58452.8 & -4.6  & Voy 2 \\
-2.4 & 60353.5 & 60260.9 &  11.1 & Voy1 \\
-9.8 & 60138.0 & 60045.4 & -2.5 & Pio 11 \\
-26.6 & 58913.4 & 58820.8 & -20.4 & Voy 2 \\
-71.2 & 54948.4 & 54855.8 & 2.7 & Voy 1 \\
\hline
\end{tabular}
\caption[Hello]{
	\label{Satradii}
	Saturn's latitude $\phi$ and radius $r$. The radii at one bar are extrapolated from the measured radii at 100 mbar under the assumption of no vertical wind gradients. The residuals are for the 100 mbar radii and are from a converged fit with the period as a free parameter. The best fit model is found for a rotation period of 10h 32m 35s. 
	}
\end{center}
\end{table}

\begin{table}
\begin{center}
\begin{tabular}{lccccl}
\multicolumn{5}{c}{} \\
\hline
\hline
Parameter & Jupiter 1 bar & Jupiter 100 mbar & Saturn 1 bar & Saturn 100 mbar\\
\hline
r$_e$ (km)  & 71,492 & 71,541 & 60,268 & 60,367 \\
r$_p$ (km) & 66,854 & 66,896 & 54,364 & 54,438 \\
GM (km$^3$ s$^{-2}$) & 126,712,765 & 126,712,765 & 37,931,207.7 & 37,931,207.7 \\
J$_2$ (10$^{-6}$) &14696.43 &14676.31 & 16324.2 & 16270.75\\
J$_4$ (10$^{-6}$) & -587.14 & -585.53 & -939.69 & -933.54\\
J$_6$ (10$^{-6}$) & 34.25 & 34.11 & 86.67 & 85.82\\
\hline
\end{tabular}
\caption[Hello]{
	\label{S}
Data are taken from JPL database:  http://ssd.jpl.nasa.gov, Jacobson (2003), Jacobson et al. (2006), and Lindal et al., (1981; 1985).
	}
\end{center}
\end{table}

\clearpage
\begin{figure}[h!]
   \centering
   \includegraphics[width=4.5in]{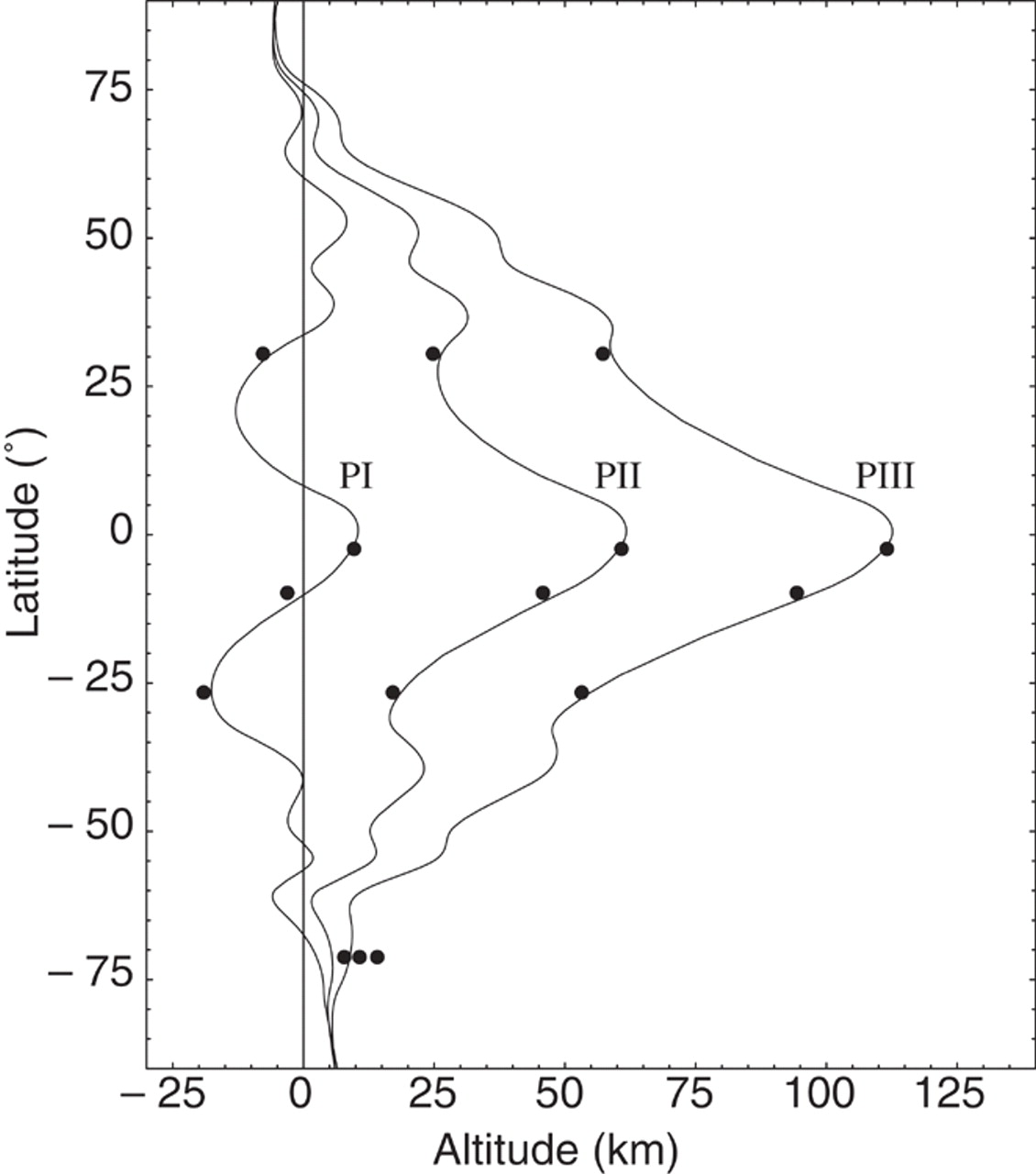}
  \caption[err]{Saturn's altitudes of the 100 mbar isobaric surface above a reference geoid with three different periods (10 hours, 32 min, 35 s (PI); 10 hours, 35min, 35 s (PII); and 10 hours, 38 min, 35 s (PIII)), and a polar radius of 54,438 km. The circles represent radii obtained by radio occultation measurements with the Pioneer 11, Voyager 1, and Voyager 2 spacecraft (Lindal et al., 1985). The solid curves represent a 100 mbar isosurface in geostrophic balance based on zonal-wind data obtained from Voyager. The vertical line represents a reference geoid unperturbed by zonal winds. }
\end{figure}

\newpage
\begin{figure}[h!]
    \centering
   \includegraphics[width=4.8in]{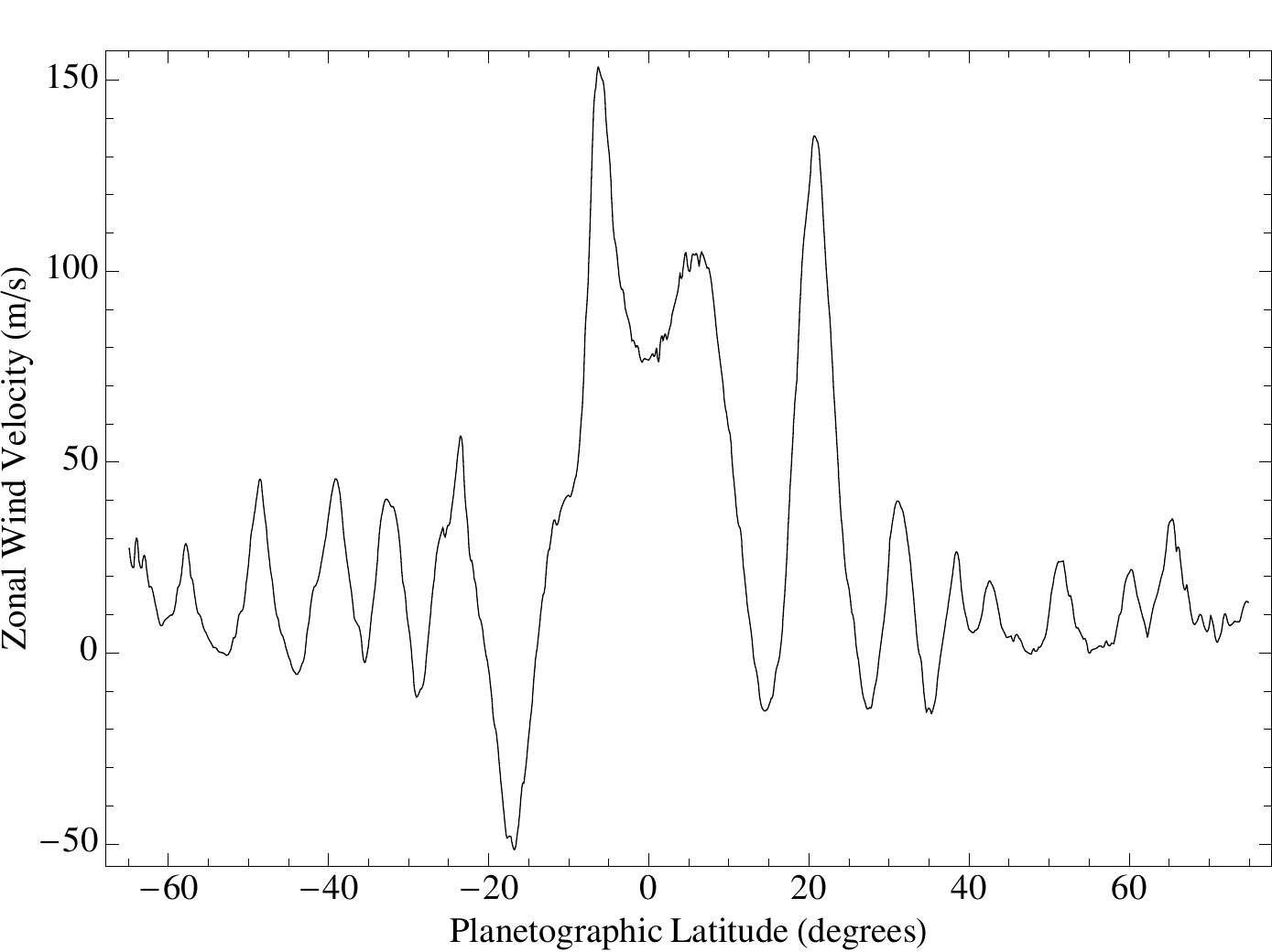}
   \caption[err]{Average zonal wind velocity profile for Jupiter, from HST images between 1995 and 1998, from Garc\'{i}a-Melendo \& S\'{a}nchez-Lavega (2001). Positive wind velocities are eastward. }
\end{figure}

\newpage
\begin{figure}[h!]
    \centering
   \includegraphics[width=4.5in]{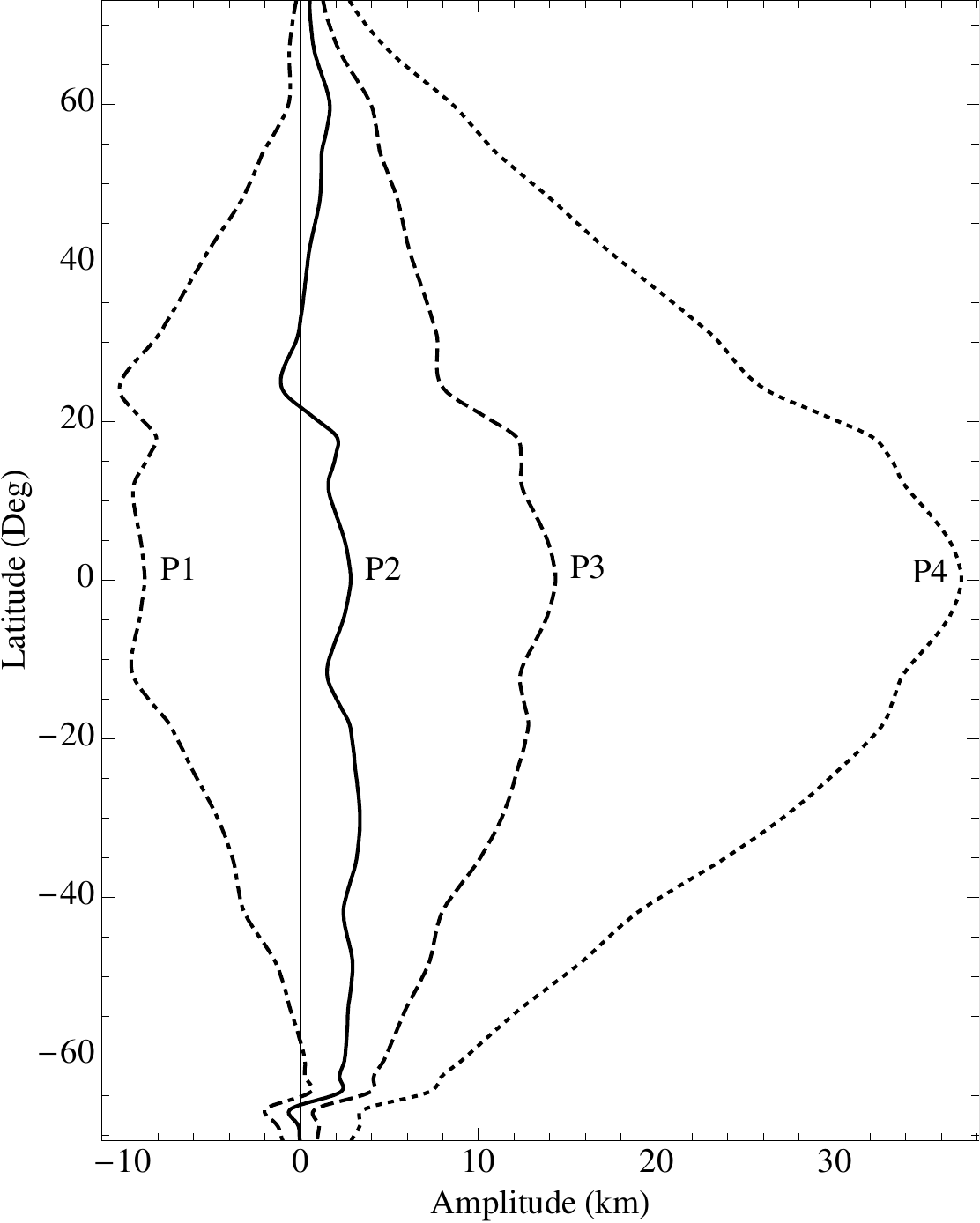}
   \caption[err]{Jupiter's altitudes of the 1 bar isosurface above a reference geoid for four different rotation periods (9 hours, 53 min, 29.7 s (P1), 9 hours, 54 min, 29.7 s (P2), 9 hours, 55 min, 29.7 s (P3), and  9 hours, 57 min, 29.7 s (P4)), and a polar radius of 66,854 km. 
The vertical line represents a reference geoid unperturbed by zonal winds. }
\end{figure}

\end{document}